\documentclass[12pt, a4paper]{article}

\newcommand{\sect}[1]{ \section{#1} }

\newcommand{\ve}{\left( \begin{array}{r}}
\newcommand{\ev}{\end{array} \right)}
\newcommand{\ar}{\left( \begin{array}{rr}}
\newcommand{\ra}{\end{array} \right)}
\newcommand{\arr}{\left( \begin{array}{rrrr}}
\newcommand{\arrr}{\left( \begin{array}{rrrrrr}}

\newcommand{\eqr}{\begin{eqnarray}}
\newcommand{\rqe}{\end{eqnarray}}
\newcommand{\eq}{\begin{equation}}
\newcommand{\qe}{\end{equation}}

\def\KK{{\rm I\kern -.2em  K}}
\def\NN{{\rm I\kern -.16em N}}
\def\RR{{\rm I\kern -.2em  R}}

\def\ZZZ{{\small{\rm Z}\kern -.5em Z}}
\def\QQ{{\rm \kern .25em
             \vrule height1.4ex depth-.12ex width.06em\kern-.31em Q}}
\def\CC{{\rm \kern .25em
             \vrule height1.4ex depth-.12ex width.06em\kern-.31em C}}

\title{${\cal N}=1$ M5-brane geometries}
\author{Bj\"orn Brinne$^1$\footnote{email: brinne@physto.se} \and
Ansar Fayyazuddin$^1$\footnote{email: ansar@physto.se} \and
Tasneem Zehra Husain$^1$\footnote{email: tasneem@physto.se} \and
Douglas J. Smith$^2$\footnote{email: Douglas.Smith@durham.ac.uk} }

\begin{document}

\maketitle

\begin{center}

{\em
\vspace{-0.5cm}
$^1$Department of Physics\\
Stockholm University\\
Box 6730\\
S-113 85 Stockholm\\
Sweden \\
\vspace{0.5cm}
$^2$Department of Mathematical Sciences \\
University of Durham \\
Durham \\
DH1 3LE \\
UK
}

\end{center}

\vspace{0.5cm}

\begin{abstract}
We describe how to construct solutions to 11-dimensional supergravity
corresponding to M5-branes
wrapped on holomorphic 2-cycles embedded in $\CC^3$. These solutions preserve
${\cal N}=1$ supersymmetry in four dimensions. In the near-horizon limit they are expected to be dual to ${\cal N}=1$ large $N$
gauge theories in four dimensions by Maldacena's duality.    
\end{abstract}

\vspace{-18.5cm}
\begin{flushright}
USITP-00-19 \\
DTP/00/111 \\
hep-th/0012194
\end{flushright}

\thispagestyle{empty}

\newpage

\setcounter{page}{1}

\sect{Introduction}
Maldacena's AdS/CFT conjecture \cite{Maldacena, AdS_refs} relates gauge theories realised as world-volume field 
theories on branes to supergravity in the near-horizon geometry produced 
by those branes. Since it is known how to describe large classes of
supersymmetric field theories using brane configurations, it is of interest to
find the corresponding near-horizon supergravity solutions.
In this paper we investigate 
the geometry produced by M5-branes wrapped on Riemann surfaces embedded in
$\CC^3$.
Particular choices of Riemann surfaces
describe intersecting M5-branes. Finding the supergravity description of fully
localised intersecting branes is a topic of current interest
\cite{intM5, danda, Rajaraman2, Youm2, loc2, loc1, loc3, loc3.5, loc4, 
loc5, loc6, pema, loc7, Gauntlett, Itzhaki, Ohta, Ohta2, ett}.
However, our main motivation is to find supergravity duals of a class of
${\cal N}=1$
gauge theories since the latter share many features
with non-supersymmetric QCD.
Finding the supergravity solutions, at least in the
near horizon limit, should provide a tool to understanding the strong coupling
limit of these field theories.

Alternative
methods for constructing supergravity duals of field theories,
such as the near-horizon limit of D3-branes in a non-trivial 
type IIB background or as solutions of five-dimensional gauged supergravity,
are described in \cite{AGMOO}.

Recently progress has been made in understanding the supergravity
description of wrapped branes in various contexts. The closely related
 ${\cal N}=2$ preserving
backgrounds describing wrapped M5-branes on 2-cycles in
$\CC^2$ were found in \cite{M5_Sigma}, and the corresponding
problem for NS5-branes in type IIA theory was studied in \cite{Ohta2}.  
M5-branes
wrapping 3-cycles of G2 holonomy manifolds were considered in the 
recent paper\cite{Acharya}, D$p$-branes wrapping collapsed cycles in 
orbifold compactifications were described in \cite{Paolo, Rajaraman},
and branes wrapping Special Lagrangian  
submanifolds were treated in \cite{Nieder}. 
Many other related examples of warped
compactifications or branes wrapped on supersymmetric cycles have been
considered in \cite{Cho, Oz, Cvetic1, Cvetic2, Cvetic3, Nunez, MaldNun}.

Some features of a large class of
supersymmetric `compactifications' (on non-compact manifolds) of eleven-dimensional supergravity were
discussed in \cite{becker}. The supergravity solutions of an M5-brane wrapped
on a Riemann surface embedded in $\CC^2$ or $\CC^3$ are special cases of these compactifications.  In \cite{becker} a number of constraints on 
supersymmetry preserving geometries are provided which our explicit
solutions satisfy.

The outline of this paper is as follows.
We generalise to ${\cal N}=1$ the construction in 
\cite{danda, intM5, M5_Sigma},
which describes the eleven-dimensional supergravity duals of 
${\cal N} = 2$ field theories as the near-horizon limit of an 
M5-brane wrapped on a Riemann
surface $\Sigma \subset \CC^2$.  
On general grounds it is expected that the supergravity
description will be valid in the large $N$ limit. The number of colours $N$ is
related to the genus of $\Sigma$ but more work needs to be done to establish
the precise region of validity of the supergravity duals.
As we will review in section 
\ref{BraneSetup}, the eleven-dimensional supergravity dual of certain 
${\cal N} = 1$ field theories (so-called MQCD theories which are in the same
universality class as supersymmetric Yang-Mills theories) is given by the
near-horizon limit of 
an M5-brane wrapped on a Riemann surface
$\Sigma \subset \CC^3$. Again we postpone the consideration of the precise
conditions for the supergravity solution to be a good description of the
field theory and concentrate on first providing a method to find explicit
solutions to the supergravity equations of motion.
We perform the first steps of this task in section 3 where we explain how to
construct the
supergravity solution by solving the appropriate BPS conditions.

\section{M5-brane setup}
\label{BraneSetup}

Brane configurations describing ${\cal N}=1$ field theories, which are
generalisations of the ${\cal N}=2$ configurations of \cite{vafa,HW,witn2},
have been constructed in \cite{EGK, EGKRS, HOO, witn1}.  The world-volume
theories were dubbed 'MQCD' to distinguish them from QCD which does
not share all the properties of the world-volume theories.

The idea is to begin with a system of NS5-branes and D4-branes
in type IIA string theory. In the simplest case there are two NS5-branes,
denoted by NS5$_1$
and NS5$_2$.  The NS5$_1$ brane has world-volume directions
$012345$, while the NS5$_2$ brane has world-volume
directions $012389$.  The NS5$_1$ and NS5$_2$ branes are separated
in the $6$ direction and we define the NS5$_1$ brane to be to the left of the
NS5$_2$ brane. One can now consider D4-branes of three distinct
types, all with world-volume directions $01236$.  We have $n_L$ D4-branes with
semi-infinite extent
in the $6$ direction ending on the left of the NS5$_1$ brane and similarly $n_R$
semi-infinite D4-branes ending on the right
of the NS5$_2$ brane. Finally $n$ D4-branes of finite extent in the $6$ direction are
suspended between the NS5$_1$ and NS5$_2$ branes. This configuration describes
an ${\cal N}=1$ four-dimensional SU($n$) field theory on the worldvolume of
the $n$ finite D4-branes. The semi-infinite branes contribute matter in both the
fundamental and anti-fundamental representations of the gauge group. So there
are $n_L + n_R$ flavours and the theory is non-chiral.

More generally the NS$5_2$ brane can be be partially rotated (or sheared
\cite{GiveonPelc}) from
the $45$ plane into the $89$ plane. This can be seen as a deformation of the
${\cal N} = 2$ theory (with the two NS5-branes parallel) by turning on a mass
for the adjoint scalar in the ${\cal N} = 2$ vector multiplet, breaking the
supersymmetry to ${\cal N} = 1$. As in the ${\cal N} = 2$ configurations, more
general setups can be considered by adding more NS5-branes with D4-branes
between them, describing an ${\cal N} = 1$ field theory with gauge group
$\prod_i {\mathrm SU}(n_i)$ and bi-fundamental matter. The NS5-branes can be
rotated by different amounts between the $45$ and $89$ planes, although
arbitrary rotations do not preserve supersymmetry (see for instance
\cite{GivKut}).  These
configurations can be lifted to M-theory where they become an M5-brane wrapped
on a non-compact Riemann surface $\Sigma$ embedded in $\CC^3$, generalising the
${\cal N} = 2$ case of $\Sigma \subset \CC^2$.
 
Further generalisations are possible, introducing D6-branes or orientifold
planes which we will not explicitly consider in this paper. These
generalisations allow the description of field theories with orthogonal and
symplectic gauge groups and more general matter contents including chiral
theories. There have been
many papers investigating properties of these brane configurations and their
relevance to field theory. We simply refer the reader to the general review of
brane configurations describing field theories in various dimensions
\cite{GivKut}.

\section{Supersymmetry preserving solutions}

In this section we present supersymmetry preserving solutions
of 11-dimensional supergravity relevant for describing the M5-brane set-up
described above.  Our starting point is a metric, generalising that
of \cite{danda}, with the isometries of the brane configuration:

\eq
ds^2= H_1^2\eta_{\mu\nu}dx^{\mu}dx^{\nu} + 2G_{M\overline{N}}dz^{M}
dz^{\overline{N}} + H_{2}^2dy^2. \label{metric}
\qe
Here the $x^\mu$ are coordinates in four-dimensional Minkowski space and $\eta_{\mu\nu}$
is the flat Minkowski metric.  The $z^M$ are holomorphic coordinates:
\eqr
z^1 & = &v = x^4 + ix^5 \nonumber \\
z^2 & = &w = x^6 + ix^7 \\
z^3 & = &s = x^8 + ix^9. \nonumber
\rqe
Four-dimensional Lorentz symmetry requires that the metric not depend
explicitly on the $x^\mu$: thus $H_1, H_2, G_{M\overline{N}}$
depend solely on $z^M, z^{\overline{N}}$ and $y$.  We are interested
in holomorphic curves in the six-dimensional space spanned by
the $z^M$, motivating the use of complex coordinates.  In addition
we have assumed that the metric in this sub-space is hermitian.

The strategy is as follows.
We are interested in finding supersymmetric solutions of eleven-dimensional supergravity.
The bosonic fields of the theory are the metric and a 3-form potential,
while the gravitino is the only fermionic field in the theory.
We set the gravitino field to zero and look for supersymmetry
preserving solutions by setting the variation of the gravitino under
supersymmetry to zero.  This fixes the 4-form field strength of the 
3-form gauge potential
in terms of the metric appearing in (\ref{metric}) as in \cite{danda}.

The gravitino variation is:
\eq
\delta\Psi_{i} = D_{i}\epsilon + \frac{1}{144}{\Gamma_{i}}^{jklm}F_{jklm}
\epsilon
-\frac{1}{18}\Gamma^{jkl}F_{ijkl}\epsilon. \label{susy}
\qe
The supersymmetry variation parameter, $\epsilon$, for our brane configuration
was determined in \cite{ansmich} where the supersymmetry properties of
precisely this configuration were investigated using the methods of
\cite{BBS}\footnote{See also \cite{ansmich1}.}.  In that work it
was established that the brane configuration is invariant under the variation
parameter with the following properties:
\eqr
\epsilon & = &\alpha + \beta \label{epsilon}\nonumber\\
\beta &=& B\alpha^* \\
i\hat{\Gamma}_0\hat{\Gamma}_1\hat{\Gamma}_2\hat{\Gamma}_3
\alpha &=& \alpha \nonumber\\
i\hat{\Gamma}_0\hat{\Gamma}_1\hat{\Gamma}_2\hat{\Gamma}_3 \beta &=& \beta
\nonumber \\
\hat{\Gamma}_{\overline{m}n}\alpha &=& \eta_{\overline{m}n}\alpha \nonumber \\
\hat{\Gamma}_{\overline{m}n}\beta &=& -\eta_{\overline{m}n}\beta \nonumber
\rqe
$\eta_{\overline{m}n}$ is the flat Euclidean metric, hatted gamma matrices
satisfy the flat space Clifford algebra, and $B$ is the charge conjugation
matrix:
\eq
\hat{\Gamma}^* = B\hat{\Gamma} B.
\qe
To proceed we set (\ref{susy}) to zero with the ansatz for the metric
(\ref{metric}) making use of the properties of $\epsilon$ given in
(\ref{epsilon}).

Setting the supersymmetry variation in (\ref{susy}) to zero results in a set
of equations relating the space-time fields to
each other. These relations can be reduced to the following set of independent
equations:
\eqr
\partial_{y}\ln{H_1}& = &-\frac{1}{12}\partial_{y}\ln{\det G} \\
\partial_{y}\ln{H_2}& = & \frac{1}{6}\partial_{y}\ln{\det G} \\
\partial_{\overline{Q}}\ln H_{2} & = &-2\partial_{\overline{Q}}H_{1} \\
F_{M\overline{Q}\overline{L}y} & = & \frac{1}{2}
\{\partial_{\overline{L}}(H_2G_{M\overline{Q}})-
\partial_{\overline{Q}}(H_2G_{M\overline{L}})\} \\
G^{M\overline{P}}G^{N\overline{Q}}F_{MN\overline{P}\overline{Q}}
&=&-\frac{1}{2}H_{2}^{-1}\partial_{y}\ln\det G\\
F_{M\overline{N}\overline{P}\overline{Q}} &=&0\\
F_{y\overline{N}\overline{P}\overline{Q}} &=&0 \\
\partial_{\overline{L}}(H_{2}^{2}G^{M\overline{L}}) & = & 0
\label{constraint}
\rqe
In addition there are equations involving derivatives of the variation
parameters which, after using the above equations, can be reduced
to:
\eqr
\partial_{i}\alpha &+& \frac{1}{4}\partial_{i}\ln H_2  = 0\nonumber\\
\partial_{\mu}\alpha & = &0
\rqe
where $i$ can be $y, M, \overline{M}$ and $\mu$ is an index in
the $0123$ directions. We have only listed independent relations,
the remaining follow from requiring that the four-form $F$ is real,
and that $\alpha$ is related to $\beta$ through (\ref{epsilon}).

Taking into account the observations of \cite{loc4, Cho},
\eq
H^2|f(z)|^2\equiv H_2^{6}|f(z)|^2 = |\det G|
\qe
with $f$ a holomorphic function of $z^{M}$.  The arbitrariness
of the function $f$ allows for the freedom to make a holomorphic
change of variables in $z^M$\cite{Cho}. In the following we will
use coordinates where $f(z) = 1$.

In addition to the supersymmetry conditions there are
constraints on the fields arising from the Bianchi identity
and equation of motion of the four-form field strength.
Since we are considering a geometry produced by M5-branes, which couple
magnetically to the three-form potential,
the roles of the Bianchi identity and equation of motion are
reversed.  Therefore we require that $d*F=0$ trivially. This
determines $F_{MN\overline{P}\overline{Q}}$ in terms of the metric:
\eq
F_{MN\overline{P}\overline{Q}} = \frac{1}{2}\partial_y[H_{2}^{-1}
(G_{N\overline{P}}G_{M\overline{Q}}-G_{N\overline{Q}}G_{M\overline{P}})].
\qe
Finally, we write down the source equation for $F$:
\eqr
J_{ML\overline{KN}y} = (dF)_{ML\overline{KN}y}  &=&
[\partial_M\partial_{\overline{N}}(H_2 G_{L\overline{K}}) -
\partial_M\partial_{\overline{K}}(H_2 G_{L\overline{N}}) \nonumber \\
&-& \partial_L\partial_{\overline{N}}(H_2 G_{M\overline{K}}) +
\partial_L\partial_{\overline{K}}(H_2 G_{M\overline{N}})] \nonumber \\
&-& \frac{1}{2}\partial_{y}^2\{H_2^{-1} (G_{M\overline{K}}
G_{L\overline{N}} -G_{L\overline{K}}G_{M\overline{N}})\}
\rqe
where $J$ is the source 5-form specifying the Riemann surface on
which the M5-brane is wrapped.  The other components of $dF$ vanish
when the constraint on the metric (\ref{constraint}) is taken into account.

The solution is expressed in a more elegant form in terms of the
rescaled metric:
\eq
g_{M\overline{N}}=H^{-\frac{1}{6}}G_{M\overline{N}},
\qe
and its associated hermitian 2-form:
\eq
\omega = ig_{M\overline{N}}dz^{M} \wedge dz^{\overline{N}}.
\qe
In terms of the rescaled metric:
\eqr
ds^{2} &=& H^{-\frac{1}{3}} \eta_{\mu\nu} dx^{\mu} dx^{\nu}
+ 2H^{\frac{1}{6}} g_{M\overline{N}} dz^{M} dz^{\overline{N}}
+H^{\frac{2}{3}} dy^{2},\nonumber \\
\det g &=& H \nonumber \\
F&=&\partial_{y}(\omega \wedge \omega)  -i\partial
(H^{\frac{1}{2}}\omega)\wedge dy
+i \overline{\partial}(H^{\frac{1}{2}}\omega)\wedge dy,\\
\overline{\partial}(\omega \wedge \omega) &=&0\nonumber.
\rqe
In the above equations $\partial$ denotes the (1,0) exterior derivative
$\partial = dz^M\partial_M$ in the subspace spanned by the $z^M$s.
Notice that the constraint on the metric (\ref{constraint})
is transformed into the property that the 4-form $\omega^2$ is closed.
Finally, the equation of motion for $F$ is written simply as:
\eq
dF = \partial_{y}^2(\omega \wedge \omega)\wedge dy -2i
\overline{\partial}\partial
(H^{\frac{1}{2}}\omega)\wedge dy = J, \label{source}
\qe
where $J$ again denotes the source 5-form.

As a consistency check one can easily see that
the ${\cal N}=2$ \cite{danda} solution satisfies the above
constraints. Another check is provided by the recent work
of Becker and Becker\cite{becker} who found constraints on
supersymmetric M-theory backgrounds with four-dimensional Lorentz
invariance and four-form flux.  Our solution
satisfies their equations, which is a non-trivial test.

\section{Conclusions and discussion}

In this paper we have found supersymmetry
preserving solutions of 11-dimensional supergravity involving
M5-branes wrapping 2-cycles in $\CC^3$.  The supergravity fields
are expressed in terms of an auxiliary hermitian metric and its
associated two-form.  The two-form satisfies a constraint
as well as a source equation.

The main motivation for studying wrapped M5-brane configurations comes
from Maldacena's conjecture relating supergravity in
near-horizon black hole geometries to Quantum Field Theories.
${\cal N}=1$ gauge theories are the most realistic supersymmetric
theories in that they display many of the features evident in ordinary QCD
such as confinement and chiral symmetry breaking.
Recent progress in identifying supergravity duals for
${\cal N}=1$ theories in IIB \cite{PolStrass, KlebStrass, 
KlebTsey, MaldNun} and M-theory \cite{Acharya2, AMV}
has resulted in new ways of understanding field theory phenomena.

The source term in equation (\ref{source}) specifies the cycle
on which the M5-brane is wrapped.  This, in turn, determines
the dual gauge theory.  To complete the program of finding
geometries dual to interesting gauge theories, we must solve the
source equation for appropriate two-cycles.  In this paper
we do not attempt to solve the source equation, but we are
heartened by the success of our attempts at solving such
equations in the ${\cal N}=2$ context \cite{M5_Sigma}.  In that work the
key observation which made finding the solution possible, was
that locally one can treat the M5-brane as
being flat.  Then the local geometry in terms of appropriate
variables is that of an ordinary flat M5-brane.  We believe
that these observations will play an important role in solving
the ${\cal N}=1$ case as well. One of the crucial simplifications in the
${\cal N} = 2$ case was to directly find the solution in the near-horizon
limit, rather than finding the full asymptotically flat solution first. We
expect this will also simplify the procedure in the ${\cal N} = 1$ case and
of course the near-horizon solution is all we are interested in finding for
the purposes of find the supergravity duals. We hope to return soon to the
question of finding explicit solutions dual
to interesting field theories.

Another motivation for studying wrapped M5-brane configurations
comes from the Randall-Sundrum \cite{RS} approach to solving the
hierarchy problem.  There, warped geometries of the kind
we display play a crucial role.  Moreover, recent results \cite{Nunez}
indicate that string theory realizations of the Randall-Sundrum
scenario miss important physics unless the full 10- or 11-dimensional
geometry is taken into account.  This problem arises due to the appearance
of singularities in geometries of lower dimensional truncated supergravity,
even in situations in which the 10- or 11-dimensional geometry is
singularity free.  Thus finding interesting ${\cal N}=1$
geometries relevant for the Randall-Sundrum scenario 
may be of phenomenological interest.

\section{Acknowledgements}
AF would like to thank Subir Mukhopadhyay for discussions.
DJS would like to thank the Department of Physics at Stockholm University and
also the members of the Department of Theoretical Physics at Uppsala University
for their hospitality.  AF is supported by a grant from the Swedish
Research Council (NFR).


\begin{thebibliography}{77}
\bibitem{Maldacena}
J.M.~Maldacena {\it The Large N Limit of Superconformal Field
Theories and
Supergravity} Adv.\ Theor.\ Math.\ Phys.\ {\bf 2} (1998) 231-252, hep-th/9711200
\bibitem{AdS_refs}
S.~Gubser, I.~Klebanov and A.~Polyakov {\it Gauge theory correlators
from
noncritical string theory} Phys.\ Lett.\ {\bf B428} (1998) 105,
hep-th/9802109;
E.~Witten {\it Anti-de Sitter space and holography} Adv.\ Theor.\ Math.\
Phys.\
{\bf 2} (1998) 253, hep-th/9802150.
\bibitem{intM5}
A.~Fayyazuddin and D.J.~Smith {\it Warped AdS near-horizon geometry of
completely localized intersections of M5-branes}
JHEP {\bf 0010} (2000) 023,
hep-th/0006060.
\bibitem{danda}
A.~Fayyazuddin and D.J.~Smith, {\it Localized
intersections of M5-branes and four-dimensional superconformal
field theories.}, JHEP {\bf 9904} (1999) 030, e-print archive:
hep-th/9902210.
\bibitem{Rajaraman2}
A.~Rajaraman,
{\it Supergravity solutions for localised brane intersections}
hep-th/0007241.
\bibitem{Youm2}
D.~Youm,
{\it Probing partially localized supergravity background of fundamental string ending on Dp-brane}
hep-th/9906232.
\bibitem{loc2} A. Hashimoto, {\it Supergravity solutions for
localized
intersections of branes.}, JHEP {\bf 9901} (1999) 018,
e-print archive: hep-th/9812159.
\bibitem{loc1}H. Yang, {\it Localized intersecting brane
solutions of D=11 supergravity.}, hep-th/9902128.
\bibitem{loc3}D. Youm, {\it Localized intersecting BPS branes.},
e-print archive: hep-th/9902208.
\bibitem{loc3.5}A. Loewy, {\it Semi Localized Brane Intersections in
SUGRA.},
Phys. Lett. {\bf B463} {1999} 41, e-print archive: hep-th/9903038.
\bibitem{loc4}A. Gomberoff, D. Kastor, D. Marolf, J. Traschen, {\it
Fully Localized Brane Intersections - The Plot Thickens},
Phys.Rev. {\bf D61} (2000) 024012, e-print archive: hep-th/9905094.
\bibitem{loc5}D. Youm, {\it Supergravity Solutions for BI Dyons.},
Phys.Rev. {\bf D60} (1999) 105006, e-print archive: hep-th/9905155.
\bibitem{loc6}S. A. Cherkis, {\it  Supergravity Solution for M5-brane
Intersection.}, e-print archive: hep-th/9906203.
\bibitem{pema}D. Marolf and A. Peet, {\it Brane Baldness vs.
Superselection Sectors.}, Phys.Rev. {\bf D60} (1999) 105007, hep-th/9903213;
A. W. Peet, {\it Baldness/delocalization in intersecting brane
systems.}, Class. Quant. Grav. {\bf 17} (2000) 1235, e-print archive:
hep-th/9910098.
\bibitem{loc7}D. Marolf and S. Surya, {\it Localized Branes and Black Holes},
Phys.Rev. {\bf D58} (1998) 124013, e-print archive: hep-th/9805121.
\bibitem{Gauntlett}
J.~Gauntlett {\it Intersecting Branes} hep-th/9705011
\bibitem{Itzhaki}
N.~Itzhaki, A.A.~Tseytlin and S.~Yankielowicz
{\it Supergravity solutions for branes localized within branes}
Phys. Lett. {\bf B432}, 298 (1998),
hep-th/9803103.
\bibitem{Ohta}
K.~Ohta and T.~Yokono,
{\it Deformation of conifold and intersecting branes}
JHEP {\bf 0002}, 023 (2000),
hep-th/9912266.
\bibitem{Ohta2}
K.~Ohta and T.~Yokono,
{\it Linear Dilaton Background and Fully Localized Intersecting Five-branes}
,hep-th/0012030.
\bibitem{ett}
J. D. Edelstein, L. Tataru and R. Tatar, 
{\it Rules for Localized Overlappings and Intersections of p-Branes },
JHEP 06 (1998) 003; hep-th/9801049. 
\bibitem{M5_Sigma}
B.~Brinne, A.~Fayyazuddin, S.~Mukhopadhyay and D.~J.~Smith,
{\it Supergravity M5-branes wrapped on Riemann surfaces and their QFT duals}
hep-th/0009047.
\bibitem{Acharya}
B.~S.~Acharya, J.~P.~Gauntlett and N.~Kim,
{\it Fivebranes wrapped on associative three-cycles}
hep-th/0011190.
\bibitem{Paolo}
M.~Bertolini, P.~Di~Vecchia, M.~Frau, A.~Lerda, R.~Marotta and I.~Pesando,
{\it Fractional D-branes and their gauge duals}
hep-th/0011077.
\bibitem{Rajaraman}
A.~Rajaraman,
{\it Supergravity duals for N = 2 gauge theories}
hep-th/0011279.
\bibitem{Nieder}
H.~Nieder and Y.~Oz,
{\it Supergravity and D-branes wrapping special Lagrangian cycles}
hep-th/0011288.
\bibitem{Cho}
H.~Cho, M.~Emam, D.~Kastor and J.~Traschen,
{\it Calibrations and Fayyazuddin-Smith spacetimes}
hep-th/0009062.
\bibitem{Oz}
Y.~Oz,
{\it Warped compactifications and AdS/CFT}
hep-th/0004009.
\bibitem{Cvetic1}
M.~Cvetic, H.~Lu, C.~N.~Pope and J.~F.~Vazquez-Poritz,
{\it AdS in warped spacetimes}
Phys.\ Rev.\  {\bf D62}, 122003 (2000),
hep-th/0005246.
\bibitem{Cvetic2}
M.~Cvetic, G.~W.~Gibbons, H.~Lu and C.~N.~Pope,
{\it Ricci-flat metrics, harmonic forms and brane resolutions}
hep-th/0012011.
\bibitem{Cvetic3}
M.~Cvetic, H.~Lu and C.~N.~Pope,
{\it Consistent warped-space Kaluza-Klein reductions, half-maximal gauged  supergravities and CP(n) constructions}
hep-th/0007109.
\bibitem{Nunez}
J.~Maldacena and C.~Nunez,
{\it Supergravity description of field theories on curved manifolds and a no go theorem}
hep-th/0007018.
\bibitem{MaldNun}
J.~M.~Maldacena and C.~Nunez,
{\it Towards the large N limit of pure N = 1 super Yang Mills}
hep-th/0008001.
\bibitem{becker}
K.~Becker and M.~Becker,
{\it Compactifying M-theory to four dimensions}
JHEP {\bf 0011} (2000) 029,
hep-th/0010282.
\bibitem{AGMOO}
O.~Aharony, S.S.~Gubser, J.~Maldacena, H.~Ooguri and Y.~Oz
{\it Large N Field Theories, String Theory and Gravity}
Phys.\ Rept.\ {\bf 323} (2000) 183,
hep-th/9905111
\bibitem{vafa} A. Klemm, W. Lerche, P. Mayr, C.Vafa, N. Warner,
{\it Self-Dual Strings and N=2 Supersymmetric Field Theory},
Nucl.\ Phys.\ {\bf B477} (1996) 746, hep-th/9604034.
\bibitem{HW}
A.~Hanany, E.~Witten {\it Type IIB Superstrings, BPS Monopoles, And
Three-Dimensional Gauge Dynamics} Nucl. Phys. {\bf B492} (1997)
152-190,
hep-th/9611230
\bibitem{witn2}
E.~Witten {\it Solutions Of Four-Dimensional Gauge Theories Via M
Theory}
Nucl. Phys. {\bf B500} (1997) 3-42, hep-th/9703166.
\bibitem{EGK}
S.~Elitzur, A.~Giveon and D.~Kutasov,
{\it Branes and N = 1 duality in string theory}
Phys.\ Lett.\ {\bf B400} (1997) 269,
hep-th/9702014.
\bibitem{EGKRS}
S.~Elitzur, A.~Giveon, D.~Kutasov, E.~Rabinovici and A.~Schwimmer,
{\it Brane dynamics and N = 1 supersymmetric gauge theory}
Nucl.\ Phys.\ {\bf B505} (1997) 202,
hep-th/9704104.
\bibitem{HOO}
K.~Hori, H.~Ooguri and Y.~Oz,
{\it Strong coupling dynamics of four-dimensional N = 1 gauge theories from  M theory fivebrane}
Adv.\ Theor.\ Math.\ Phys.\ {\bf 1} (1998) 1,
hep-th/9706082.
\bibitem{witn1}
E.~Witten,
{\it Branes and the dynamics of {QCD}}
Nucl.\ Phys.\ {\bf B507} (1997) 658,
hep-th/9706109.
\bibitem{GiveonPelc}
A.~Giveon and O.~Pelc,
{\it M theory, type IIA string and 4D N = 1 SUSY SU(N(L)) $\times$ SU(N(R)) gauge theory}
Nucl.\ Phys.\ {\bf B512} (1998) 103,
hep-th/9708168.
\bibitem{GivKut}
A.~Giveon and D.~Kutasov,
{\it Brane dynamics and gauge theory}
Rev.\ Mod.\ Phys.\ {\bf 71} (1999) 983,
hep-th/9802067.
\bibitem{ansmich}
A.~Fayyazuddin and M.~Spalinski,
{\it The Seiberg-Witten Differential From M-Theory}
Nucl.\ Phys.\ {\bf B508} (1997) 219,
hep-th/9706087.
\bibitem{BBS}
K.~Becker, M.~Becker and A.~Strominger,
{\it Fivebranes, Membranes and Non-Perturbative String Theory}
Nucl.\ Phys.\ {\bf B456} (1995) 130,
hep-th/9507158.
\bibitem{ansmich1}
A.~Fayyazuddin and M.~Spalinski,
{\it Extended Objects in MQCD}
hep-th/9711083.
\bibitem{PolStrass}
J.~Polchinski and M.J.~Strassler,
{\it The String Dual of a Confining Four-Dimensional Gauge Theory}
hep-th/0003136.
\bibitem{KlebStrass}
I.R.~Klebanov and M.J.~Strassler,
{\it Supergravity and a Confining Gauge Theory: Duality Cascades and $\chi$SB-Resolution of Naked Singularities}
JHEP {\bf 0008} (2000) 052,
hep-th/0007191.
\bibitem{KlebTsey}
I.R.~Klebanov and A.A.~Tseytlin,
{\it Gravity Duals of Supersymmetric SU(N) $\times$ SU(N+M) Gauge Theories}
Nucl.\ Phys.\  {\bf B578}, 123 (2000)
hep-th/0002159.
\bibitem{Acharya2}B. ~Acharya,
{\it On Realising N=1 Super Yang-Mills in M theory}, hep-th/0011089.
\bibitem{AMV}
M.~Atiyah, J.~Maldacena and C.~Vafa,
{\it An M-theory Flop as a Large N Duality}
hep-th/0011256.
\bibitem{RS}
L.~Randall and R.~Sundrum,
{\it An Alternative to Compactification}
Phys.\ Rev.\ Lett.\ {\bf 83} (1999) 4690,
hep-th/9906064.
\end{thebibliography}
\end{document}